\documentclass[reprint,twocolumn,amsmath,amssymb,aps,superscriptaddress]{revtex4-2}

\usepackage{graphicx}
\usepackage{dcolumn}
\usepackage{bm}
\usepackage{graphicx}
\usepackage{amssymb}
\usepackage{epstopdf, epsfig}
\usepackage{yfonts}
\usepackage[colorlinks=true, linkcolor=blue, citecolor=blue, urlcolor=blue]{hyperref}
\usepackage{fontenc}
\usepackage{upgreek}
\usepackage{mathrsfs}
\usepackage{booktabs}
\usepackage{pifont}
\usepackage{amsmath}
\usepackage{bm}
\usepackage{wasysym}
\usepackage{caption}
\usepackage{subcaption}
\usepackage{euscript}
\usepackage{natbib}
\usepackage{lipsum}
\usepackage{xcolor}

\usepackage{orcidlink}

\usepackage{subcaption}
\usepackage{ragged2e}
\DeclareCaptionJustification{justified}{\justifying}
\newcommand{\jfm}[1]{\textcolor{blue}{#1}}
\newcommand{\x}[1]{\textcolor{black}{#1}}

\begin{document}

\title{Direction Symmetry of Wave Field Modulation by Tidal Current}

\author{Ina Teutsch\,\orcidlink{0000-0001-8751-8096}}
\email{ina.teutsch@baw.de}
\affiliation{Helmholtz-Zentrum Hereon, Coastal Climate and Regional Sea Level Changes, Max-Planck-Stra{\ss}e 1, 21502 Geesthacht, Germany}
\affiliation{Federal Waterways Engineering and Research Institute (BAW), Hamburg, 22559, Germany}
\author{Saulo Mendes\,\orcidlink{0000-0003-2395-781X}}
\email{saulo.dasilvamendes@unige.ch}
\affiliation{Group of Applied Physics, University of Geneva, Rue de l'\'{E}cole de M\'{e}decine 20, 1205 Geneva, Switzerland}
\affiliation{Institute for Environmental Sciences, University of Geneva, Boulevard Carl-Vogt 66, 1205 Geneva, Switzerland}
\affiliation{\x{University of Michigan–Shanghai Jiao Tong University Joint Institute, Shanghai Jiao Tong University, Shanghai 200240, China}}
\author{Jérôme Kasparian\,\orcidlink{0000-0003-2398-3882}}
\email{jerome.kasparian@unige.ch}
\affiliation{Group of Applied Physics, University of Geneva, Rue de l'\'{E}cole de M\'{e}decine 20, 1205 Geneva, Switzerland}
\affiliation{Institute for Environmental Sciences, University of Geneva, Boulevard Carl-Vogt 66, 1205 Geneva, Switzerland}

\begin{abstract}
Theoretical studies on the modulation of unidimensional regular waves over a flat bottom due to a current typically assign an asymmetry between the effects of opposing/following streams on the evolution of major sea variables, such as significant wave height. The significant wave height is expected to monotonically increase with opposing streams and to decrease with following streams. To some extent, observations on data sets containing a few thousand of waves or over a continuous series of about a day confirm this prediction. \x{B}ased on a multi-year dataset, we show that in very broad-banded seas with high directional spread especially at high values of the ratio between tidal stream and group \x{velocity}, the asymptotic behavior of sea variables is highly non-trivial and does not follow the theoretical predictions, \x{thus featuring a symmetrical modulation of several sea state variables respective to whether the current opposes or follows the direction of wave motion}.

\end{abstract}

\keywords{Non-equilibrium statistics ; Rogue Wave ; Stokes perturbation ; Bathymetry}

\maketitle

\section{INTRODUCTION}

The deterministic study of inhomogeneous waves sprung near the middle of the twentieth century
\citep{Unna1942,Johnson1947}. Advanced mathematical techniques arose in the following decades \citep{Ursell1960,Whitham1962,Whitham1965,Bretherton1968}, in particular with wave-current interaction being assessed through ray theory \citep{Arthur1950,Whitham1960}, linear wave theory \citep{Taylor1955,Peregrine1976}, radiation stress \citep{Higgins1960,Higgins1961}, \x{action balance \citep{Willebrand1975},} spectral \citep{Huang1972} and perturbative methods \citep{McKee1974}. Through these theoretical advances, transformations of wave characteristics were measured and confirmed in major global currents, such as the Agulhas current \citep{Irvine1988}, the Gulf Stream \citep{Holthuijsen1991,Wang1994} and the Kuroshio \citep{Hwang2005,Wang2020}. These transformations concern both significant wave height and wave energy, spectral bandwidth and directional spreading. Additionally, the transformations have been corroborated in laboratory experiments \citep{Thomas1981,Thomas1990,Swan2001,MacIver2006,Ma2010}. The first studies on wave-current systems focused on tides~\citep{Unna1942,Sverdrup1944}, \x{finding} that currents either reduce or expand the wavelength, whose change is met with a modulation in the wave height \citep{Unna1942}. These properties of the \x{quasi-stationary and/or homogeneous} wave-current system\x{s} seem to be universal, featuring similar changes in wave height, length, and period in estuaries and rivers \citep{Gonzalez1984,Holthuijsen1989,Zippel2017}. \x{On the other hand, \citet{Barber1949} was one of the first to address currents as dominantly changing in time and imply that the absolute period of a near-monochromatic swell is modified following the Doppler effect due to currents, but that the wavelength remains unchanged. As pointed out by \citet{Tolman1990},} simplified quasi-stationary or quasi-homogeneous \x{theoretical} models based on the \x{action} balance equations underestimate the effect of tidal currents on wave heights. \x{Furthermore, \citet{Tolman1991} and \citet{Holthuijsen1991} studied the problem with a full non-stationary and non-homogeneous approach, showing non-local aspects of wave-current interactions in the Southern North Sea and Gulf Stream, respectively.}

\begin{figure*}
\centering
    \includegraphics[scale=0.34]{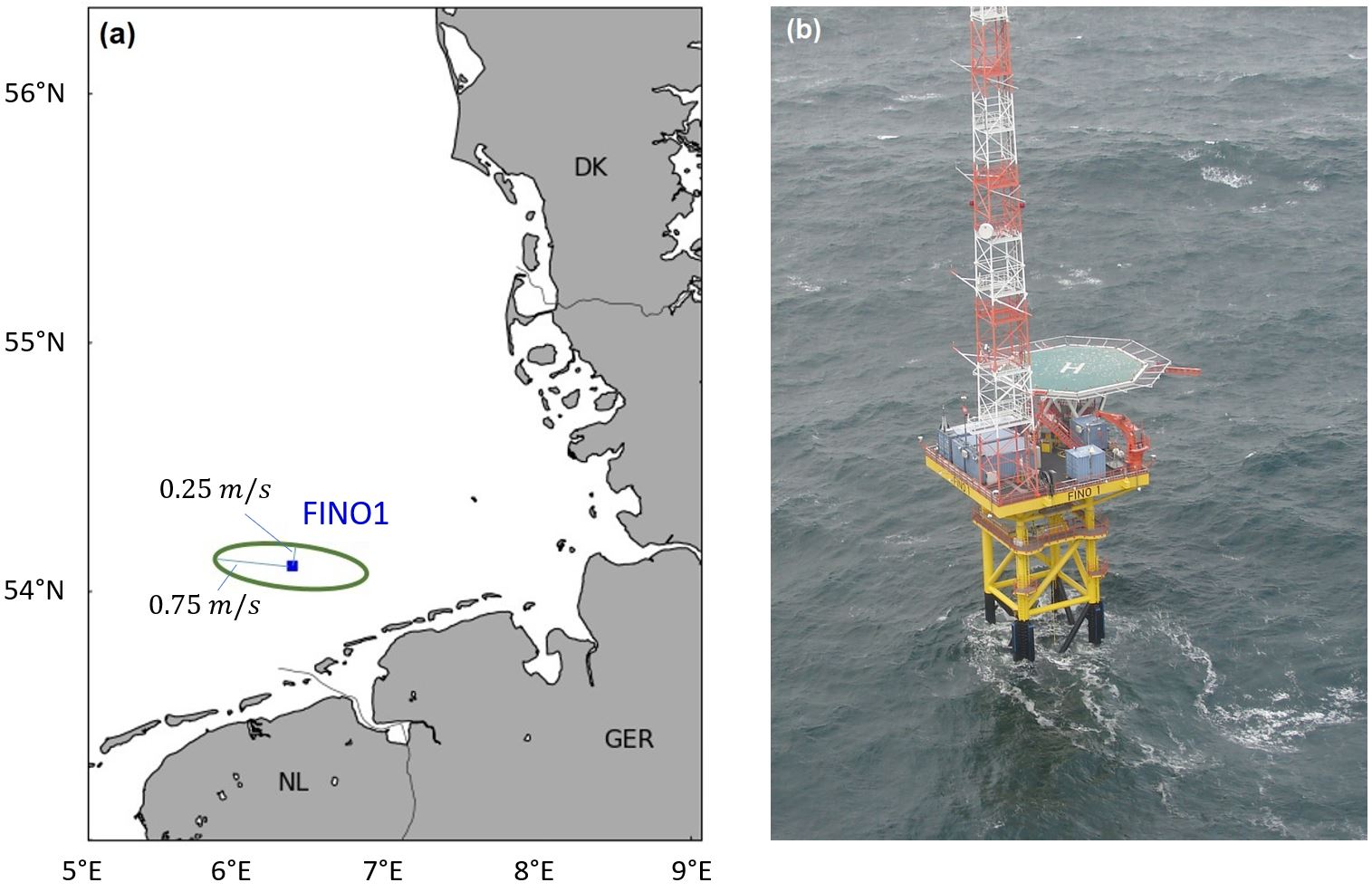}
    \caption{(a) Location of the research platform FINO1 in the southern North Sea, close to the Dutch and German Frisian islands. The green ellipse indicates the tidal cycle at the site, with semi-major and semi-minor axes of approximately 0.75 m/s and 0.25 m/s, respectively. (b) The research platform FINO1 has been constructed close to the wind farm Alpha Ventus in the North Sea to observe hydrographical and meteorological parameters, among others. \copyright Forschungs- und Entwicklungszentrum Fachhochschule Kiel GmbH}
    \label{fig:map}
\end{figure*}
The majority of the mentioned studies limit the wave-current system to waves encountering an opposing current, resulting in a shortening and steepening of waves due to energy focusing. Only a few studies considered the case of waves propagating in the same direction as the current, and their findings were ambiguous. Theoretical considerations suggested that the wavelength should increase when waves propagate in the same direction as a tidal stream \citep{Unna1942}. 
Indeed, \citet{MacIver2006} reported from an experimental study that in following currents waves become longer, while at the same time their height is reduced. They concluded that in waves with a following current, wave energy is decreased due to the strain rate of the current speed. \citet{Swan2001}, however, found in an experimental study, supported by a numerical model, that the crest-trough \x{unevenness} of waves propagating in a following current increases towards higher and sharper wave crests. 
This \x{unevenness} should increase the probability of extreme waves~\cite{Marthinsen1992,Mori1998,Mori2002b,Tayfun2020,BMendes2020,Mendes2021b,Mendes2021c}.

Observational studies to date are mostly focused on deterministic investigations for specific events or short-term time series analysis \citep{Guillou2017,Lewis2019,Barnes2020,Pizzo2023,Jia2024}, even those with larger data sets \citep{Ardhuin2009,Ardhuin2012,Gemmrich2012,Benetazzo2024}. Here, we investigate the behavior of the modulation due to the wave-current interaction of long-term series from a statistical point of view, namely of average wave properties. We characterize the symmetry in the evolution of fundamental wave properties in response to opposing and following currents \x{in the limited context of currents being co-linear to the direction of wave motion}. We show that the following currents can amplify wave fields in deep water in addition to coastal waters \citep{Pizzo2023} and that both the nonlinearity of the sea state and the speed of the tidal stream impact the magnitude and asymmetry of the modulation.

\section{Data and methods}\label{sec:intro}

\begin{figure*}
\hspace{1.5cm}
\minipage{0.38\textwidth}
    \includegraphics[scale=0.45]{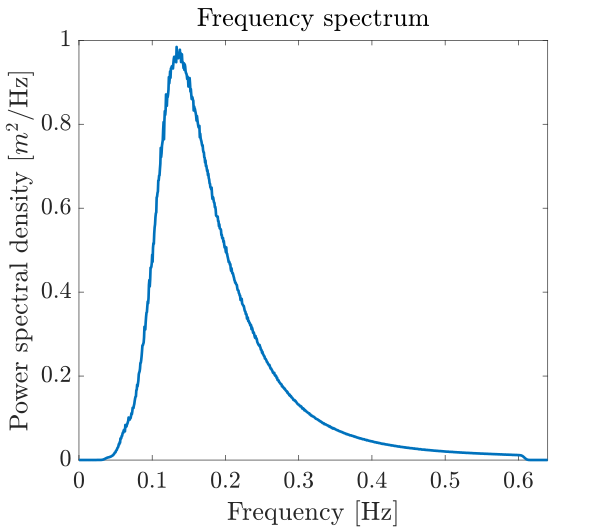}
\endminipage
\hfill
\minipage{0.49\textwidth}
    \includegraphics[scale=0.3]{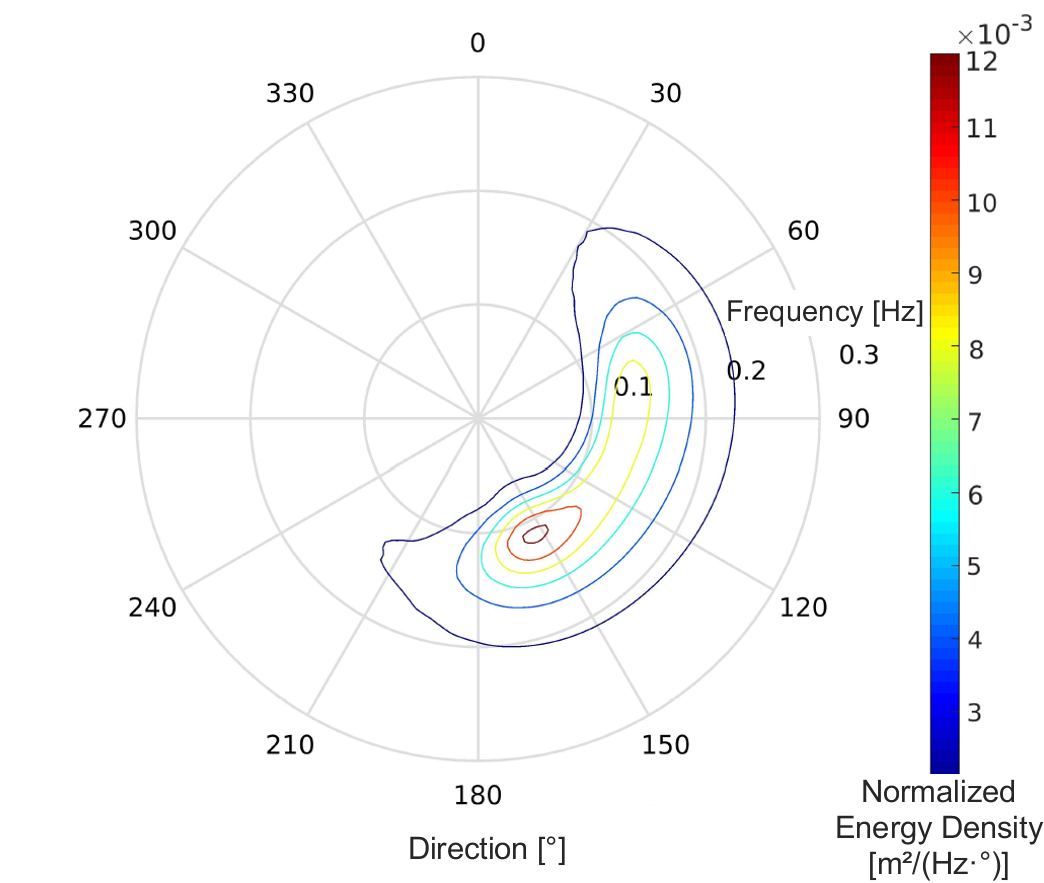}
\endminipage
\caption{Mean wave spectra during the period between 2019 and 2022 at the considered site. (a) Unfiltered one-dimensional spectrum. (b) Mean directional wave spectrum, calculated using the DIrectional WAve SPectra Toolbox \citep[DIWASP;][]{John2002}.}
    \label{fig:spectrum}
\end{figure*}

Wave and current data were recorded between July 2019 and December 2022 at about 200~m distance from the research platform FINO1 in the southern North Sea, which is located at 54.015$^{\circ}$N 6.588$^{\circ}$E in a water depth of approximately 30~m (figure~\ref{fig:map}). The currents at this site result from the tidal cycle, which forms an ellipse with a major axis stretching approximately from west-northwest to east-southeast (figure~\ref{fig:map}). Waves typically propagated either towards the southeast or towards the east, with peak wave frequencies between 0.11~Hz and 0.14~Hz (figure~\ref{fig:spectrum}). A more detailed description, of the tidal current in this region, including a tidal chart, can e.g. be found in Reynaud and Dalrymple~\citep{Reynaud2012}.
\begin{figure*}
    \centering
    \includegraphics[scale=0.18]{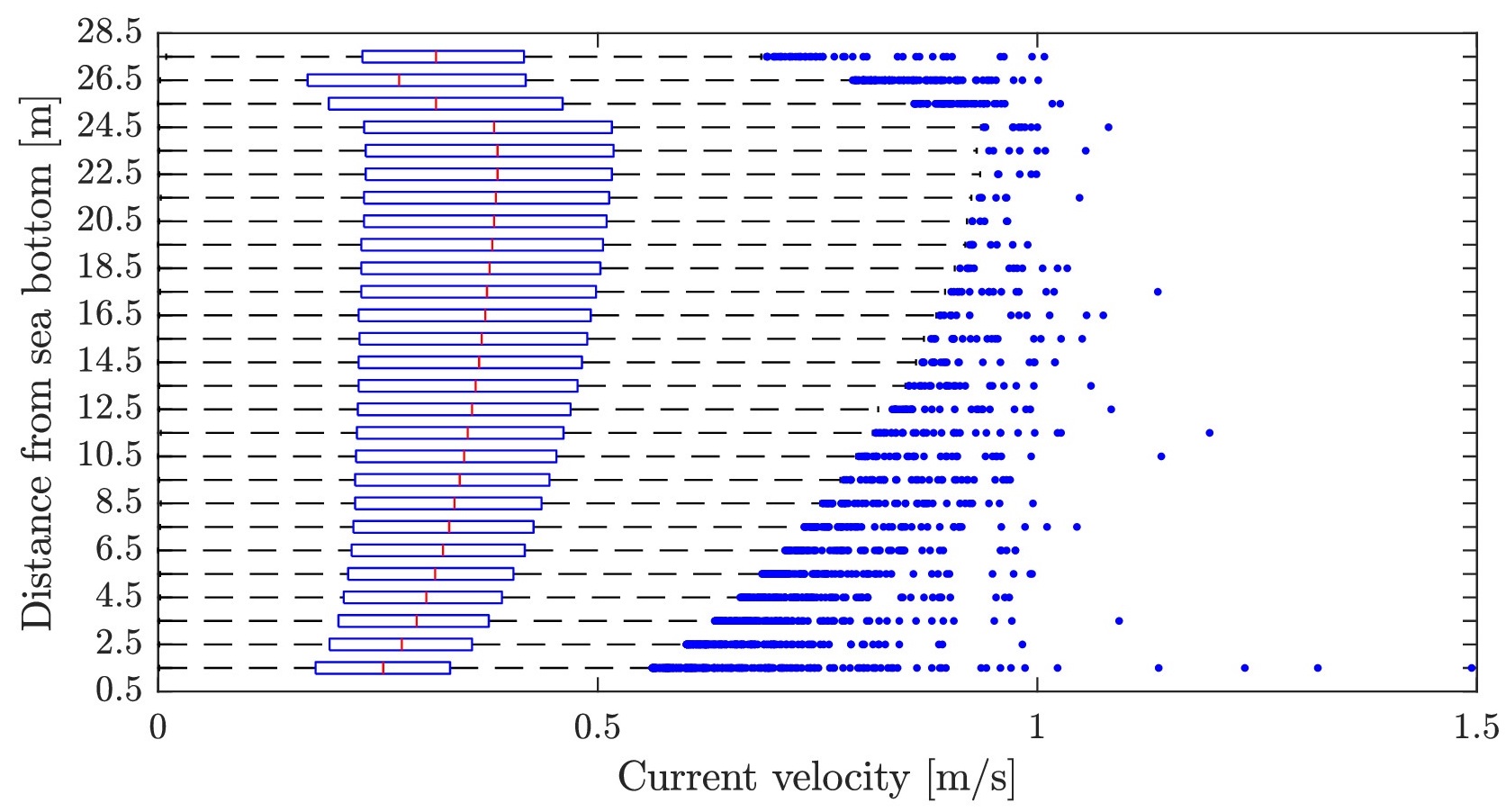}
    \caption{Box plot of approximately 40,000 measured current speed profiles during the considered time period of 3.5 years. Boxes enclose the two central quartiles with the median in red, while the wiskers encompass from the $25^{th}$ to $75^{th}$ percentiles and the blue markers indicate outliers of the distributions.}
    \label{fig:profile}
\end{figure*}
\begin{table}
  \centering
\begin{tabular}{ l| l }    
\hline
Wave  parameters & Definition\\
\hline
Peak wavelength $\lambda_\mathrm{p}$ & Solution to $(\frac{2\pi}{T_\mathrm{p}})^2=\frac{2\pi g}{\lambda_\mathrm{p}} \tanh(\frac{2\pi h}{\lambda_\mathrm{p}})$\\
Peak wave number & $k_\mathrm{p}=\frac{2\pi}{\lambda_\mathrm{p}}$\\
Relative water depth & $k_\mathrm{p}h$\\
group \x{velocity} &  $c_\mathrm{g}=\frac{1}{2} \sqrt{\frac{g}{k_\mathrm{p}}\tanh(k_\mathrm{p}h)}\cdot(1+\frac{2k_\mathrm{p}h}{\sinh(2k_\mathrm{p}h)})$\\
Spectral bandwidth & $\nu=\sqrt{\frac{m_0 m_2}{m_1^2}-1}$\\
Steepness & $\varepsilon= k_p H_s$\\
\bottomrule
\end{tabular}
\caption{Sea parameter definitions, with gravity $g$ and spectral moments $m_n$ \x{assuming small current effects}.}
\label{tab:param}
\end{table}
\begin{figure*}
\hspace{-0.8cm}
\minipage{0.3\textwidth}
    \includegraphics[scale=0.43]{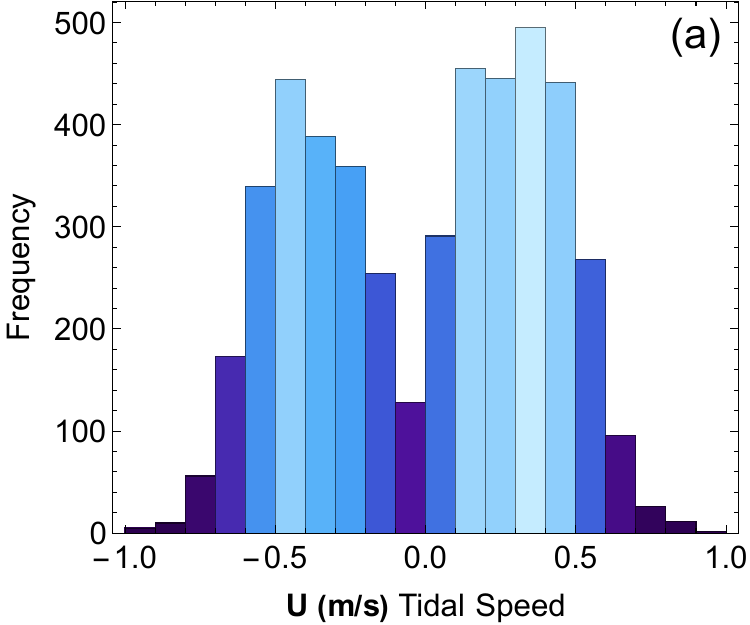}
\endminipage
\hfill
\minipage{0.29\textwidth}
    \includegraphics[scale=0.43]{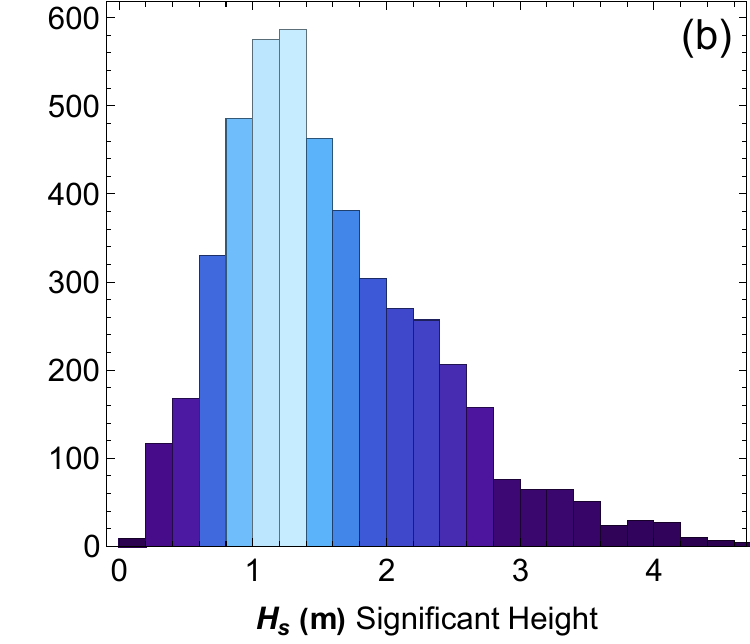}
\endminipage
\hfill
\minipage{0.35\textwidth}
    \includegraphics[scale=0.43]{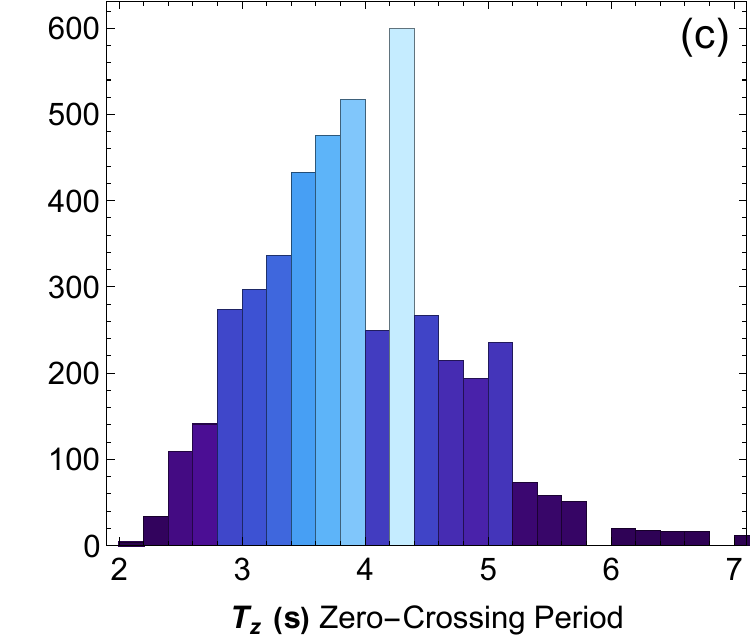}
\endminipage

\hspace{-0.7cm}
\minipage{0.3\textwidth}
    \includegraphics[scale=0.43]{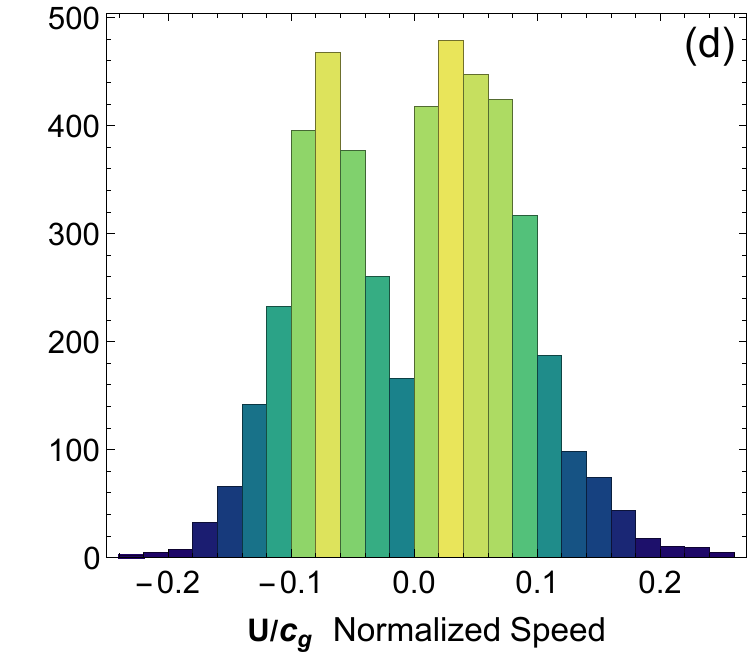}
\endminipage
\hfill
\minipage{0.28\textwidth}
    \includegraphics[scale=0.43]{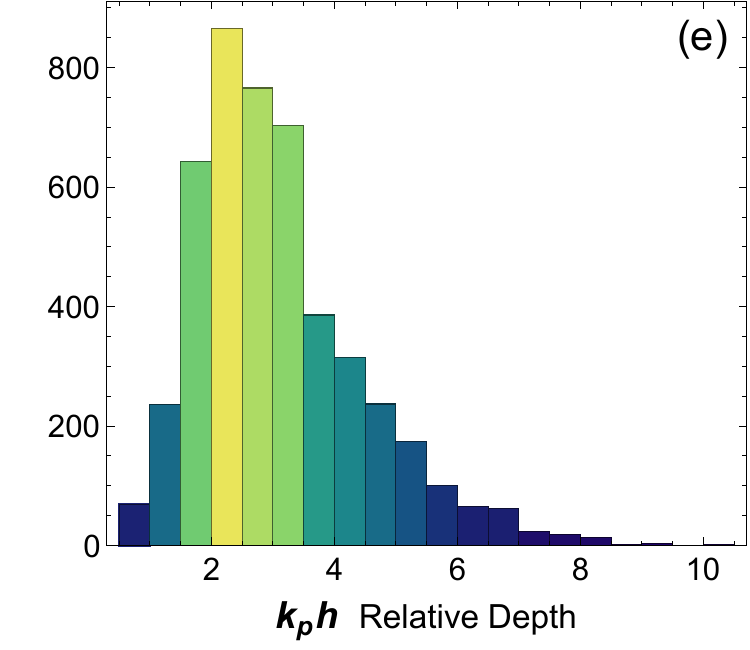}
\endminipage
\hfill
\minipage{0.35\textwidth}
    \includegraphics[scale=0.43]{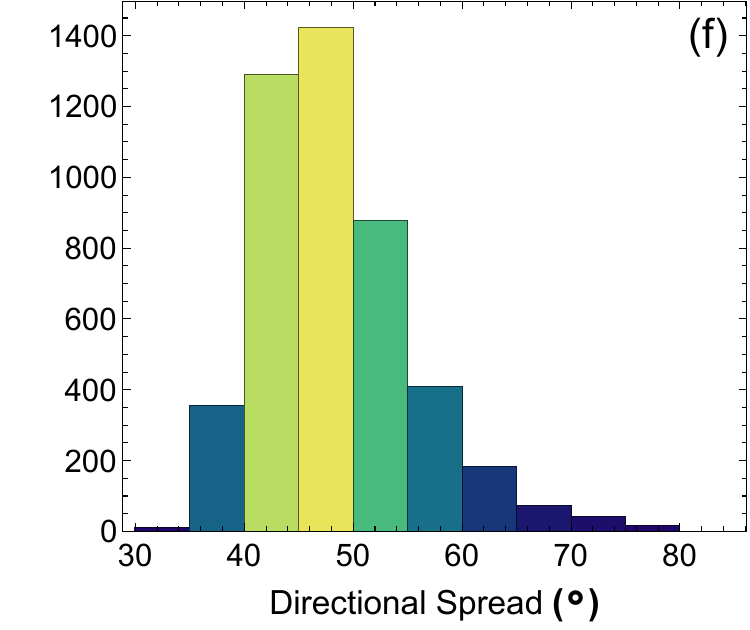}
\endminipage

\vspace{0.30cm}
\hspace{0.80cm}
\minipage{0.45\textwidth}
    \includegraphics[scale=0.43]{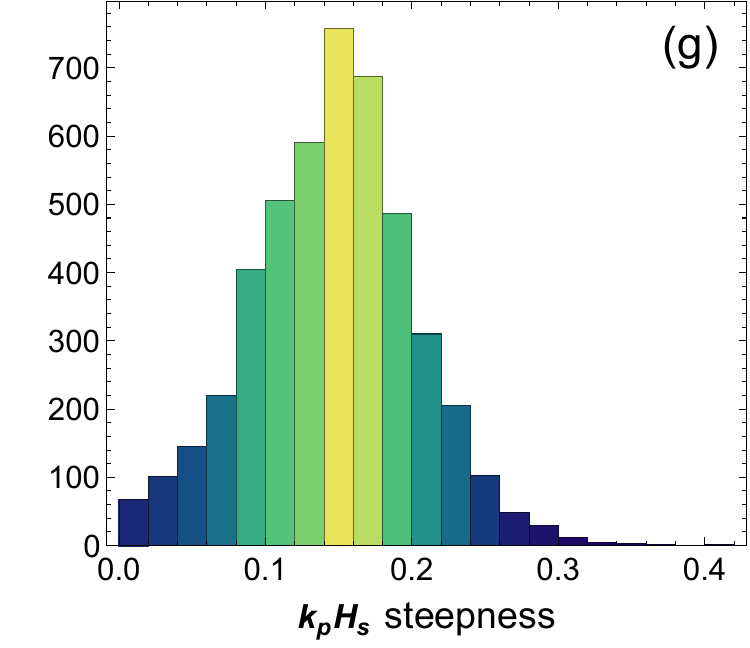}
\endminipage
\hfill
\minipage{0.5\textwidth}
    \includegraphics[scale=0.43]{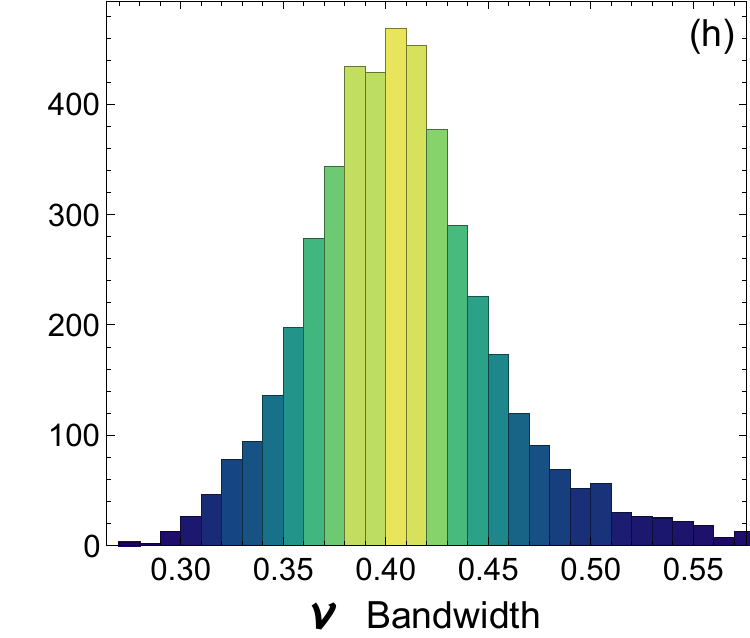}
\endminipage

\caption{Histograms of the frequency of 30-minute wave records as a function of the major dimensional variables, as well as normalized current speed. \x{Brighter colors emphasize higher probability density.}}
\label{fig:hist1}
\end{figure*}

Wave elevation data were recorded by a surface-following directional Waverider buoy of type MkIII.
The sampling frequency was 1.28~Hz. The resolution of the wave buoy is specified as 0.01 m, with an accuracy of less than  0.5\% of the measured surface elevation relative to the calibrated still water line \citep{Datawell2023}. The wave buoy delivered surface elevation data in samples of 30~minutes in length, containing 366 waves on average. The samples were quality-controlled according to the procedure described in \citet{Teutsch2020} and subsequently used to calculate the directional spectra (figure~\ref{fig:spectrum}\jfm{a,b}) and the sea state parameters in table~\ref{tab:param} \x{for the finite water depth water waves without current effects. The current effect on dispersion has ramifications to all other variables, and reads (bold denotes vectors) \citep{Higgins1960}:} 
\begin{equation}
\x{
\omega = \omega_{0} +\textbf{U}\cdot \textbf{k} \quad , \quad  \omega_{0}= g|\textbf{k}| \tanh{|\textbf{k}|h}
}
\end{equation}
\x{thus leading to,}
\begin{equation}
\x{
\omega^2_{0} = \omega^2 \left(1 - \frac{ \textbf{U}\cdot \textbf{k}}{ \omega} \right)^2 = g|\textbf{k}| \tanh{|\textbf{k}|h}
}
\end{equation}
\x{As observed in \jfm{figure} \ref{fig:hist1}, most of the data lies in the deep water regime, such that the hyperbolic function has small impact on the dispersion. Moreover, the typical values of the variables measured by their peak parameters show that $U \sim \pm 0.5$m/s, has periods of $T_p \sim 6$s and length $\lambda_p \sim 75$m and therefore the correction due to the current on the dispersion is small $\textbf{U}\cdot \textbf{k}/\omega \sim U T_p/\lambda_p \approx 0.04 \ll 1$. Therefore, we can compute peak spectral parameters without taking into account the current effect.}

Current velocities and directions were recorded by an acoustic Doppler current profiler (ADCP, Nortek), which was deployed at the sea bottom in an approximate water depth of 30~m. 
The current data made available to us by the Federal Maritime and Hydrographic Agency (BSH) were already quality-controlled using the Storm software provided by the manufacturer~\citep{Storm2022}. This included corrections for a possible tilt of the instrument and the removal of echo spikes, among others. The ADCP delivered current speeds averaged over a time window of 10~minutes. These current speeds were further time-averaged over the whole duration of each wave elevation sample. 
Samples with missing wave or current data were excluded from the analysis.

The current speed has a smooth mean vertical profile except at the boundaries (\jfm{figure} \ref{fig:profile}). Close to the bottom and the surface, wave movements, tidal elevation and sidelobe interference of the ADCP \citep{ADCP2022} perturb the measurement. Furthermore, the peak frequency of 0.12~Hz (see figure \ref{fig:spectrum}a) corresponds to wavelengths of more than 100~m, much larger than the water depth of 30~m. Waves are therefore sensitive to the current in the entire water column, although their sensitivity slightly decays with the depth. Hence, we considered the current speed approximately 6~m below the water surface (i.e., 24.5~m above the bottom, \jfm{figure} \ref{fig:profile}), as representative of the whole column. 

\begin{figure*}
\hspace{0.0cm}
\minipage{0.49\textwidth}
\hspace{0.0cm}
    \includegraphics[scale=0.42]{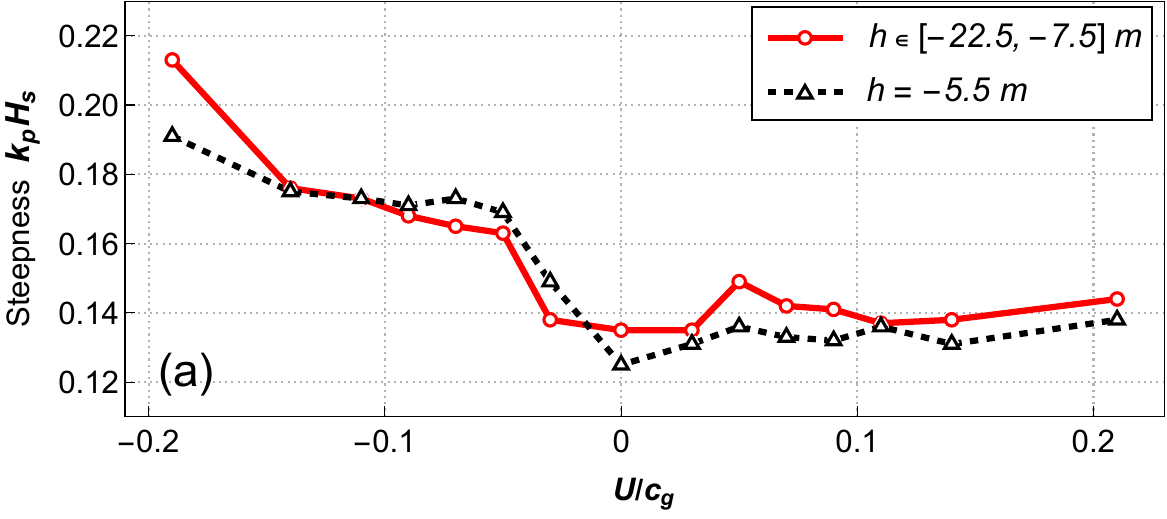}
\endminipage
\hfill
\minipage{0.49\textwidth}
\hspace{0.0cm}
    \includegraphics[scale=0.42]{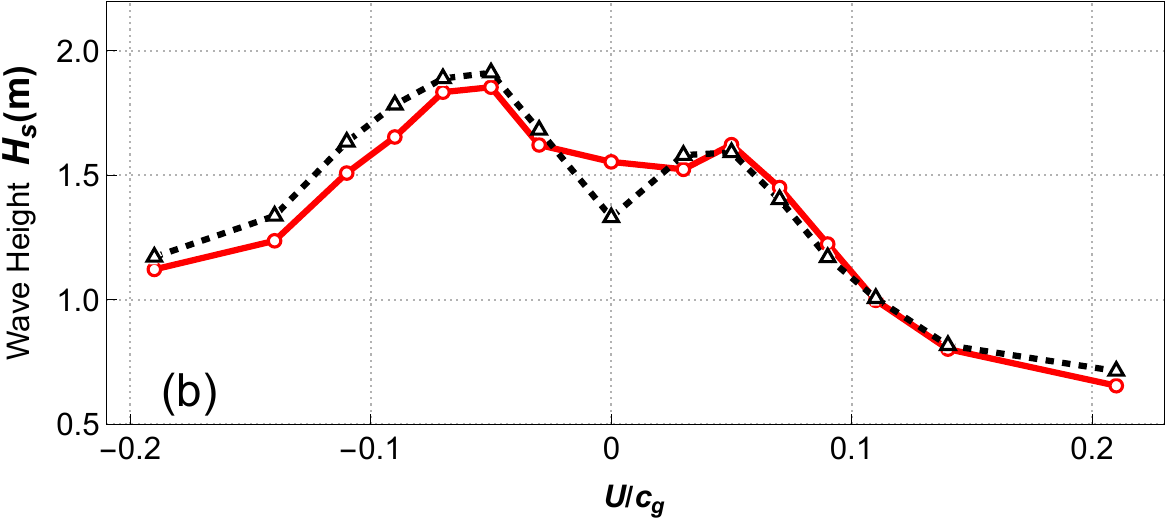}
\endminipage

\hspace{+0.5cm}
\minipage{0.4\textwidth}
    \includegraphics[scale=0.41]{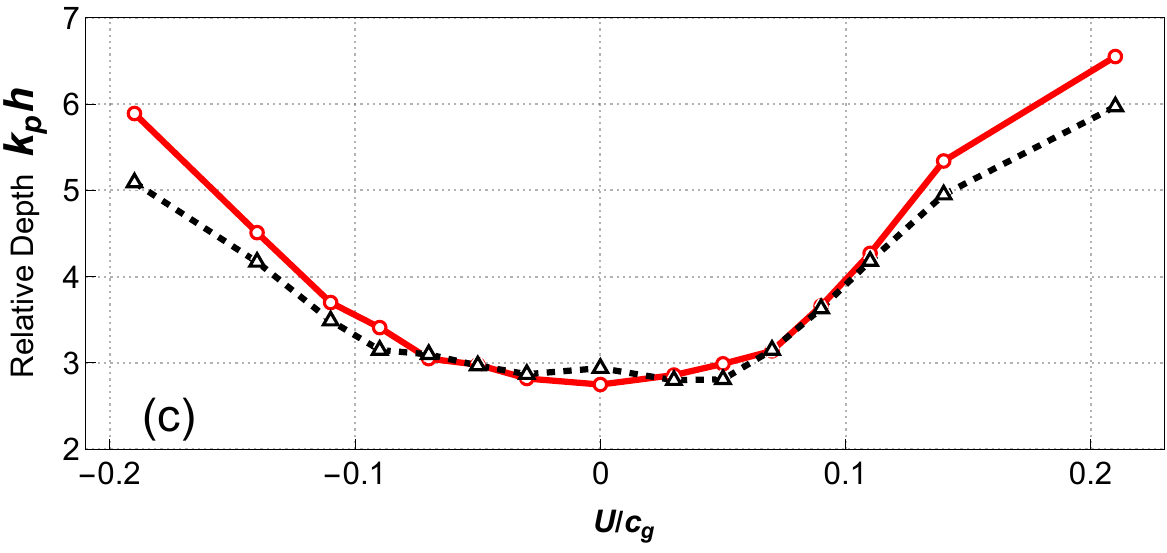}
\endminipage
\hfill
\minipage{0.49\textwidth}
    \includegraphics[scale=0.42]{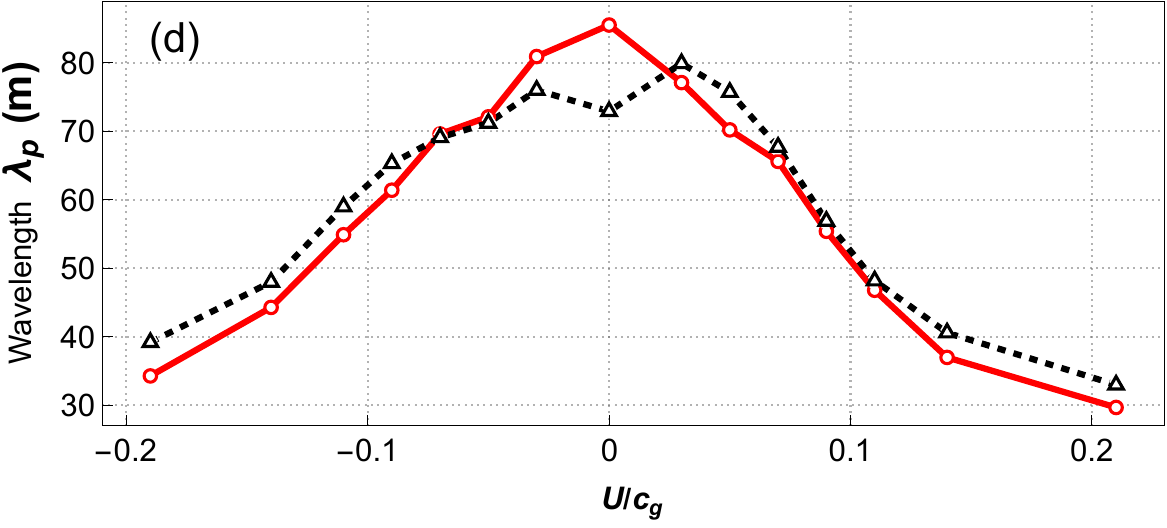}
\endminipage

\hspace{0.0cm}
\minipage{0.49\textwidth}
\hspace{0.0cm}
    \includegraphics[scale=0.42]{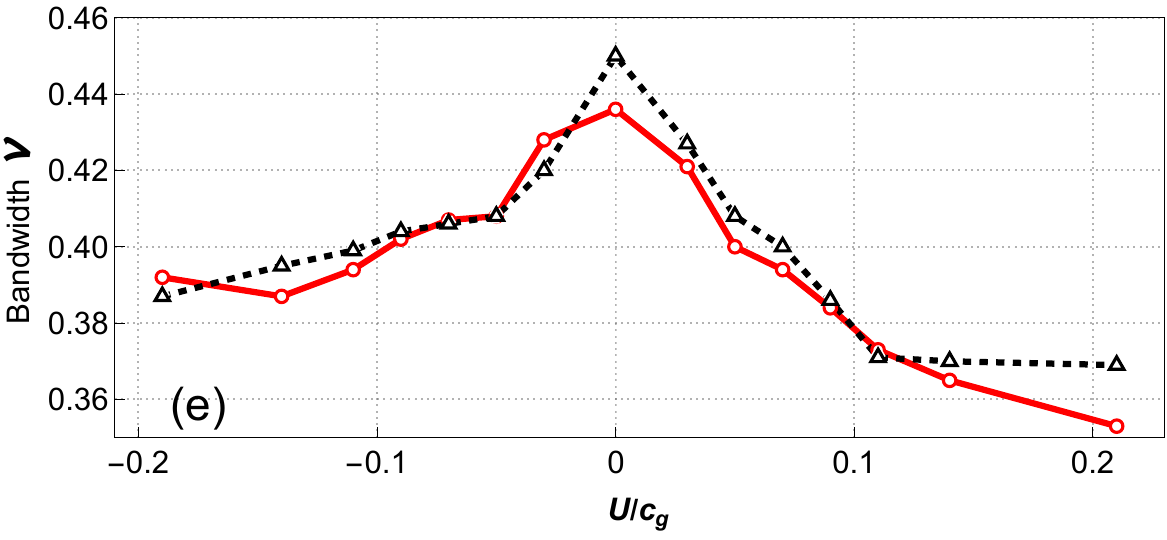}
\endminipage
\hfill
\minipage{0.49\textwidth}
\hspace{0.0cm}
    \includegraphics[scale=0.42]{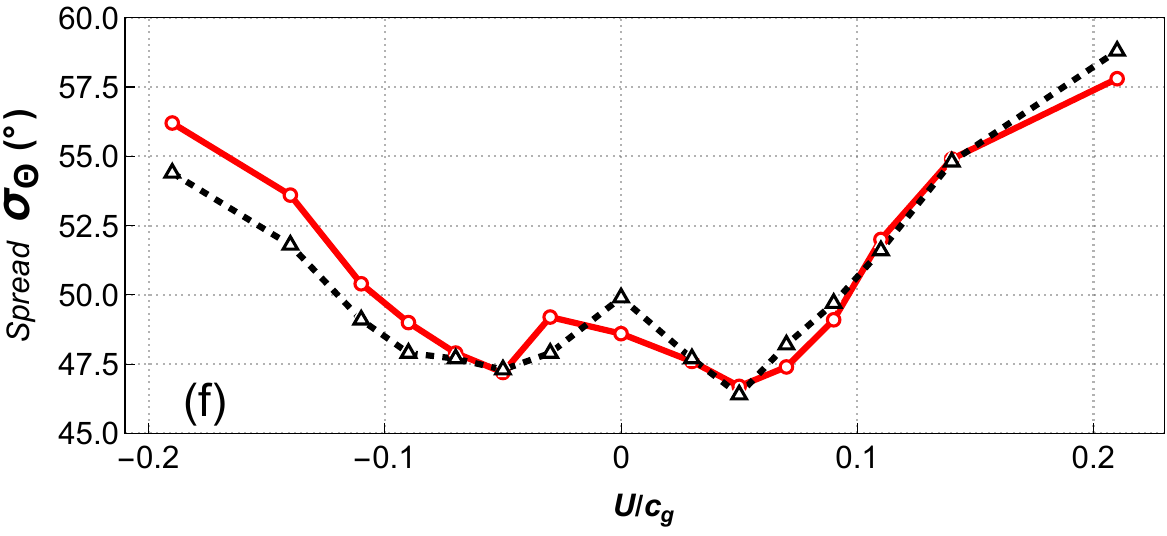}
    \endminipage
\caption{Sensitivity of results to the depth at which the current speed is measured: Average values of the sea parameters as a function of the relative current speed. The red curves are taken as a function of the normalized speed measured as an average between the 25th and 75th percentile of the water depth ($-22.5\textrm{~m} < h < -7.5\textrm{~m}$).}
\label{fig:spectraldist0}
\end{figure*}

Furthermore, current turbulence generated in the water column by the waves~\citep{Swan2001} via bottom friction~\citep{Wolf1999} may also feedback on the waves themselves. The mixed layer depth, in which turbulence may be generated by surface waves, ranges  between 2~m and 6~m in our conditions \citep[eq. 7]{Babanin2006}. Although this feedback likely averages out over the column in our measurements, we verified that neither wave-induced current turbulence, nor our choice of considering the current at 6~m depth influenced the results: repeating the analysis while considering the speed in the middle of the water column (15~m depth) or the (vector) averaged speed between 7.5~m and 22.5~m yielded similar results, as will be discussed in detail in the next section (see also \jfm{figure}~\ref{fig:spectraldist0}).

\begin{figure*}
\minipage{0.49\textwidth}
    \includegraphics[scale=0.46]{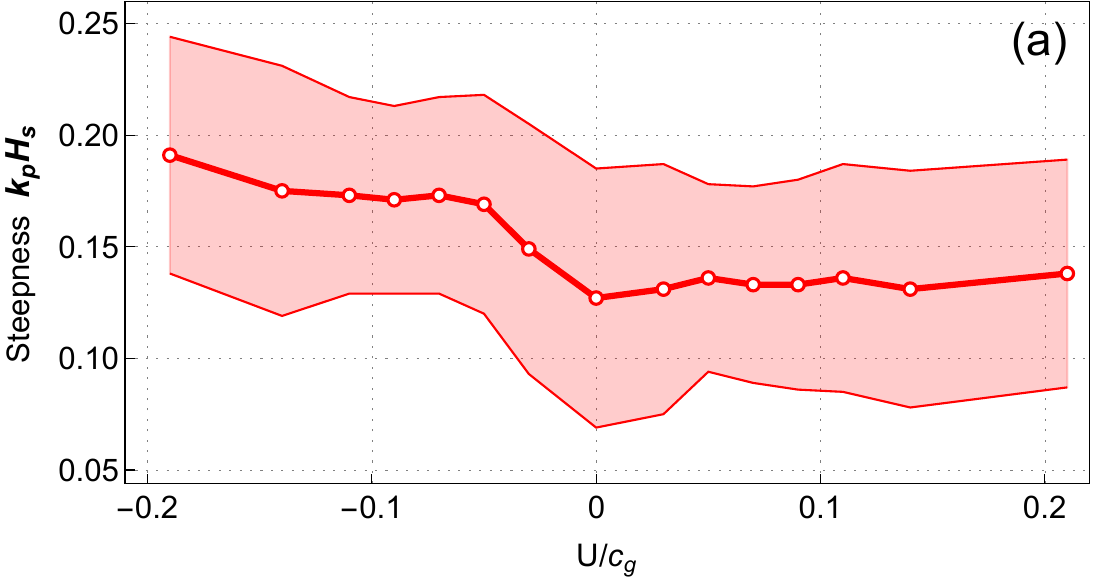}
\endminipage
\hfill
\minipage{0.49\textwidth}
\hspace{0.0cm}
    \includegraphics[scale=0.45]{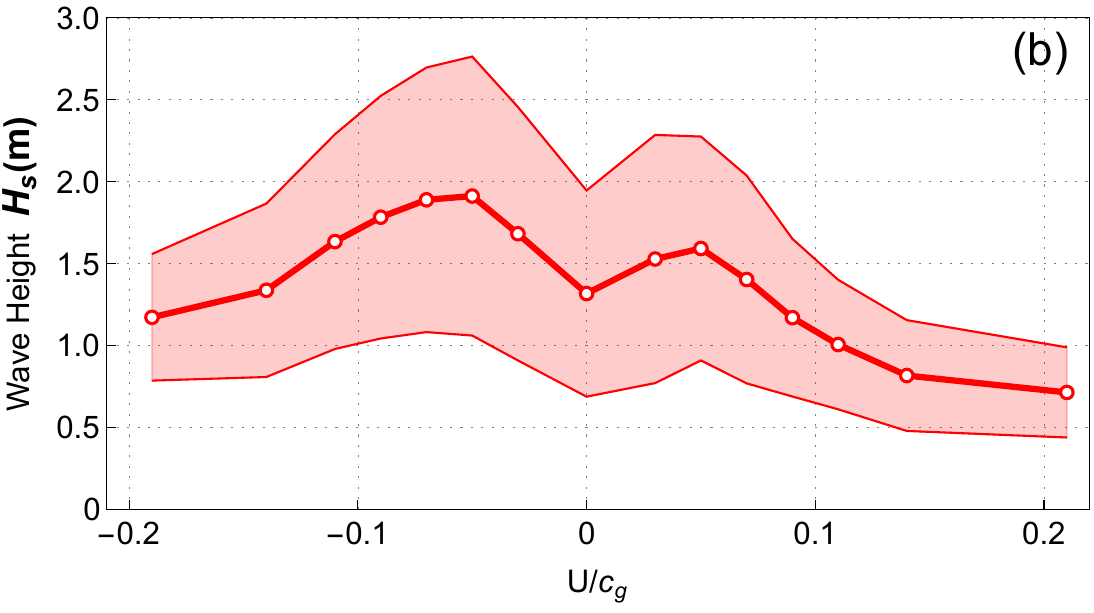}
\endminipage

\hspace{0.0cm}
\minipage{0.49\textwidth}
\hspace{0.0cm}
    \includegraphics[scale=0.44]{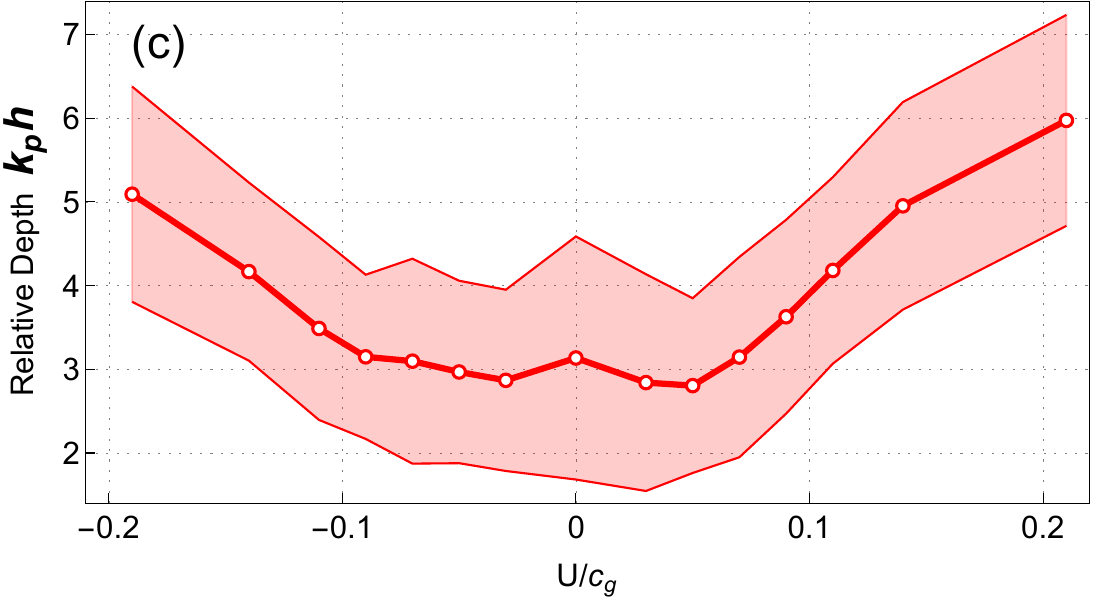}
\endminipage
\hfill
\minipage{0.49\textwidth}
\hspace{0.0cm}
    \includegraphics[scale=0.45]{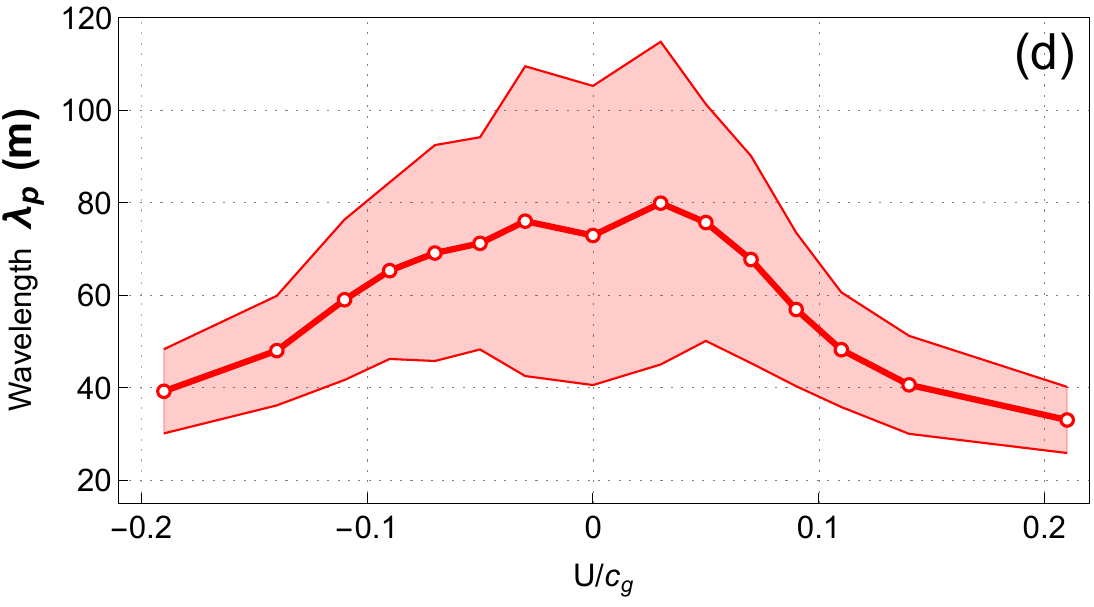}
\endminipage
\hspace{0.0cm}
\minipage{0.49\textwidth}
\hspace{0.0cm}
    \includegraphics[scale=0.45]{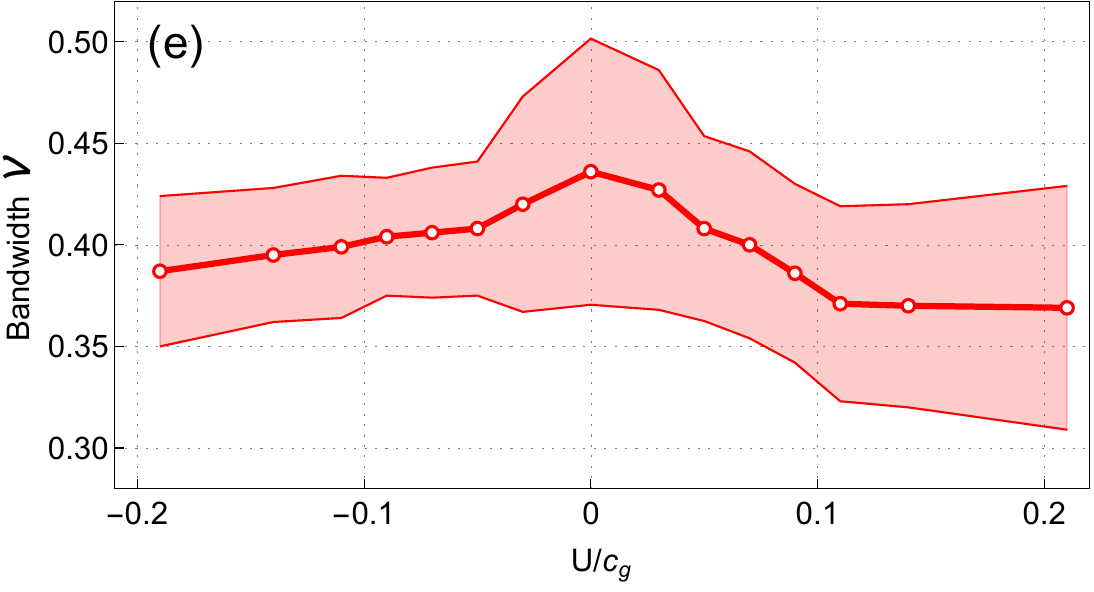}
\endminipage
\hfill
\minipage{0.49\textwidth}
\hspace{0.0cm}
    \includegraphics[scale=0.45]{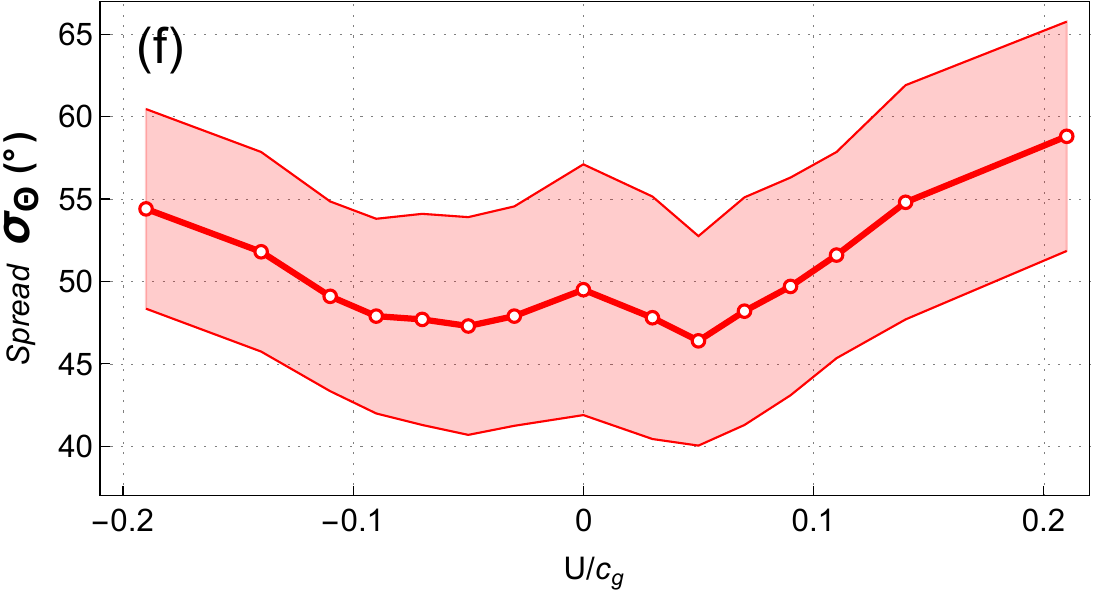}
\endminipage
\caption{Average values of the sea state parameters as a function of the relative current speed. 
Red bands depict plus or minus one standard deviation on the wave data.}
\label{fig:spectraldist00}
\end{figure*}
Wave direction is defined as the propagation direction of the waves at the peak frequency in the directional wave spectrum, $\Theta_p$. The directional spreading of the waves around the peak frequency is calculated as 
\begin{equation}
    \sigma_{\Theta_p} = \sqrt{2(1-C_1)},
\end{equation} 
with $C_1 = \sqrt{a_1^2+b_1^2}$, in which $a_1$ and $b_1$ denote the first sine and cosine Fourier coefficients of the directional spectrum, respectively \x{\citep{Kuik1988}. Note that alternative and equivalent descriptions for the directional spectrum and spreading exist, such as directional spread $\sigma_{\theta} = \sqrt{2(1+s)}$ arriving from a spectral function $D(\theta) \sim \cos^{2s}{(\theta/2)}$ \citep{Mitsuyasu1975}}. The significant wave height $H_s$ was computed as the mean of the highest third of waves in each 30-minute sample. To evaluate the effect of current speed $U$ and current direction on major sea state parameters, we selected two sets of samples, for which the angle between peak wave direction and current direction was either $0 \pm 10^{\circ}$ (henceforth referred to as "following current") or $180 \pm 10^{\circ}$ (henceforth referred to as "opposing current"). Within these ranges, sine values keep below 0.17 and cosine values stay beyond $\pm 0.98$, so that the transverse flow is negligible and the axial current is virtually unaffected. References without current, hereafter denoted rest conditions, were defined as samples with $|U/c_{g}| < 0.02$, $c_g$ being the group \x{velocity} of the peak frequency of the wave spectrum. This threshold is consistent with the residual transverse current at current incidence angles of  $\pm 10^{\circ}$.

In total, the quality-controlled data set contained 4686 samples (2156 with opposing current, 2321 with \x{following} current, 209 in rest conditions). Values of $U/c_{g}$ were binned in intervals of $\sim 1/50$, except around zero, where a single bin covers $|U/c_{g}| \leq 0.02$, and for the outermost bins, where data are sparser.

The histograms in \jfm{figure} \ref{fig:hist1} provide descriptive statistics of all selected 30~minute samples along the entire data set. \jfm{Figures} \ref{fig:hist1}\jfm{a,d} describe the near-symmetrical character of the tidal speeds, whether in dimensional units or normalized by the peak group \x{velocity}. \jfm{Figures} \ref{fig:hist1}\jfm{b,c} provide the wave climate of the region, with mild sea characteristics for both significant wave height and zero-crossing period.  Most of the data is in intermediate ($k_p h = 0.3-3.1$, 47\% of the dataset) to deep water ($k_p h > \pi$, 50\% of the dataset, or even 79\% corresponding to $k_p h \geq 2$) (\jfm{figure} \ref{fig:hist1}\jfm{e}). Adding to the complexity of the sea conditions, the spectrum is significantly directionally spread (\jfm{figure} \ref{fig:hist1}\jfm{f}). Furthermore, the spectrum is very broad (\jfm{Figure} \ref{fig:hist1}\jfm{h}). The sea is composed of waves typically of second order in wave steepness ($0.05 \leq \varepsilon \leq 0.09$) \cite{LeMehaute1976}, albeit smaller fractions of the dataset are also of first and higher orders (\jfm{figure}~\ref{fig:hist1}\jfm{g}). \x{To be precise, Stokes expansion of the velocity potential and surface elevation allows us to describe water waves even with fifth-order theory \citep{Fenton1985} regardless of the numerical value of $\varepsilon$. Nonetheless, in the range $0.05 \leq \varepsilon \leq 0.09$ the second-order theory is sufficient}. As discussed in the next section, this strong nonlinearity will be key to the physical behavior of the waves. Typically, consideration of the nonlinear regime of the wave steepness and its effect on wave-tide modulation is overlooked in most studies.

\section{Results}

\begin{figure*}
\minipage{0.6\textwidth}
    \includegraphics[scale=0.6]{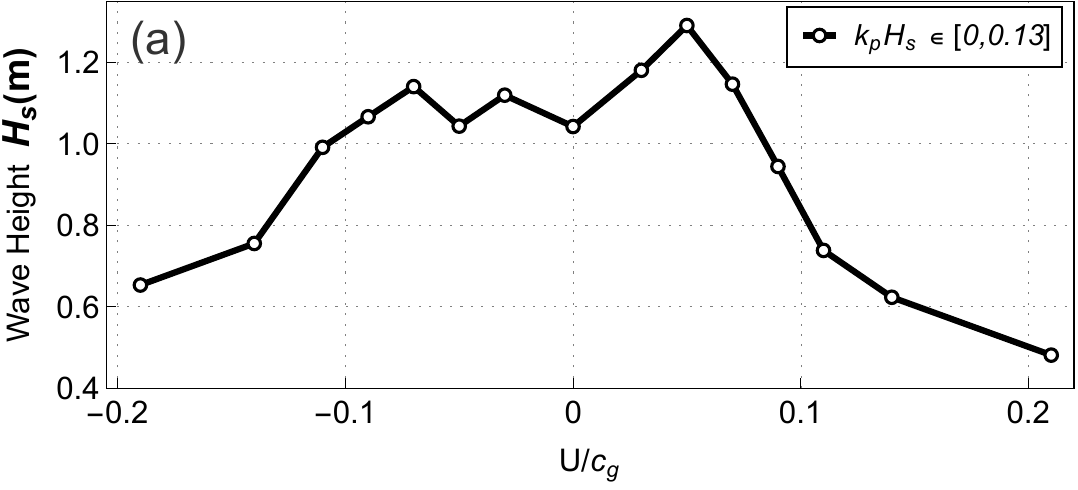}
\endminipage

\minipage{0.6\textwidth}
    \includegraphics[scale=0.6]{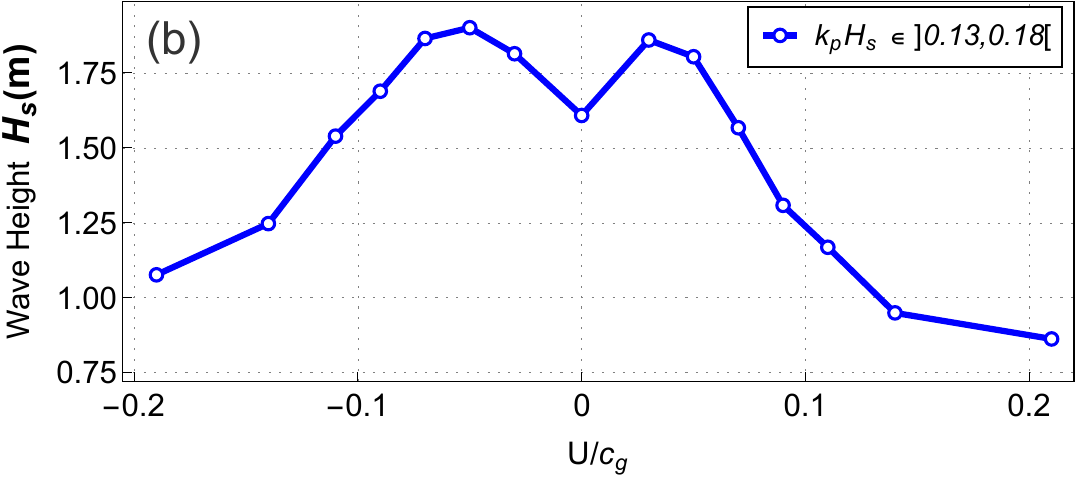}
\endminipage

\minipage{0.6\textwidth}
    \includegraphics[scale=0.6]{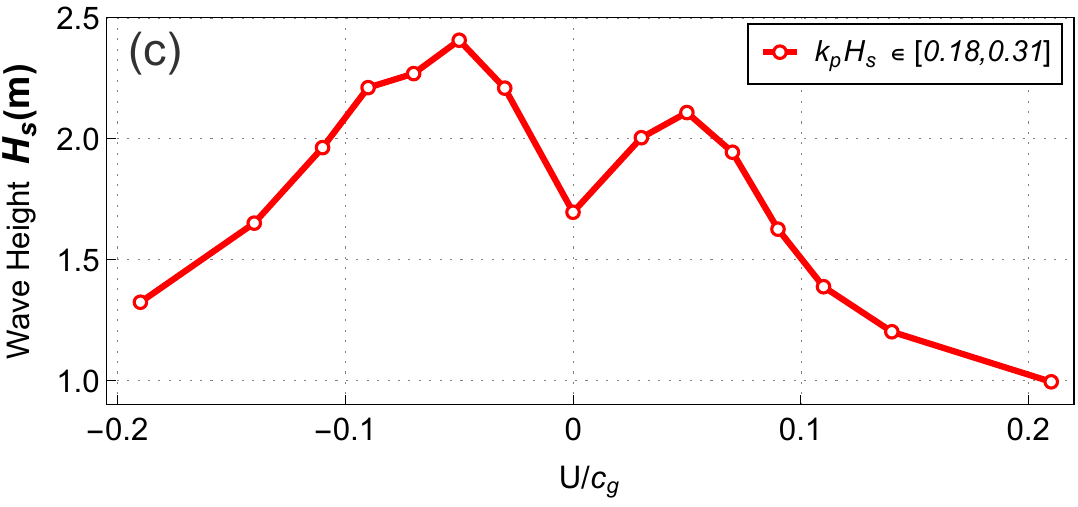}
\endminipage
\caption{Average values of the significant wave height as a function of the relative current speed with varying levels of nonlinearity.}
\label{fig:spectraldist2x}
\end{figure*}
Figure \ref{fig:spectraldist00} displays the dependence of the main wave parameters on the relative current velocity. \x{Note that there are two ways to normalize the current: either by the current-free group velocity $\bf{c_g}$ or the total group velocity $\bf{c_h} = \bf{c_g} + \bf{U}$ that is affected by the current. Following \citet{Higgins1960}, since we are dealing with data whose collected waves are co-linear to the current we can write the normalized speed as either $U/c_g = \gamma$ or alternatively $U/c_h \equiv \gamma/(1 + \gamma)$. One can verify that the modulations in amplitude and wavenumber described by \citet{Higgins1960} can all be expressed as functions of $\gamma$ without further algebra to otherwise have functions of $\gamma/(1 + \gamma)$. Thus, we proceed to plot normalized velocities in the direction of motion as $U/c_g$.} Negative current velocity refers to a current opposing the peak wave direction. Only the mean wave steepness (\jfm{figure} \ref{fig:spectraldist00}\jfm{a}) displays a clear asymmetry between forward and opposing currents. In spite of the wide absolute confidence ranges, the size of the samples is quite large, typically with 100,000 entries per bin. As such, pairwise Welch's $t$-tests~\cite{Welch1947} show a highly significant difference ($p < 10^{-3}$) between bins corresponding to forward and opposing currents of the same speed, $\pm U/c_g$, for all pairs of bins. 

Other variables display a much more symmetric behavior. 
The anticipated increase in significant wave height due to an opposing current \citep{Unna1942,Peregrine1976} is observed in our data (\jfm{figure} \ref{fig:spectraldist00}\jfm{b}), but restricted to $|U/c_{g}| \leq 0.05$. Beyond, this trend is inverted and $H_{s}$ decreases again. This decrease might be due to either wave blocking \x{\citep{Kirby2002}} or increasing dissipation due to high mean steepness, although it seems to appear earlier than expected by \citet{Toffoli2013}. The behavior is comparable to the effect of \x{following} currents: $H_s$ increases for mild current speeds, unlike the expected behavior of regular waves \citep{MacIver2006}. As in the case of an opposing stream, the significant wave height decreases for $U/c_g > 0.05$, likely due to wave dissipation. 

\begin{figure*}
\hspace{-0.5cm}
\minipage{0.6\textwidth}
    \includegraphics[scale=0.53]{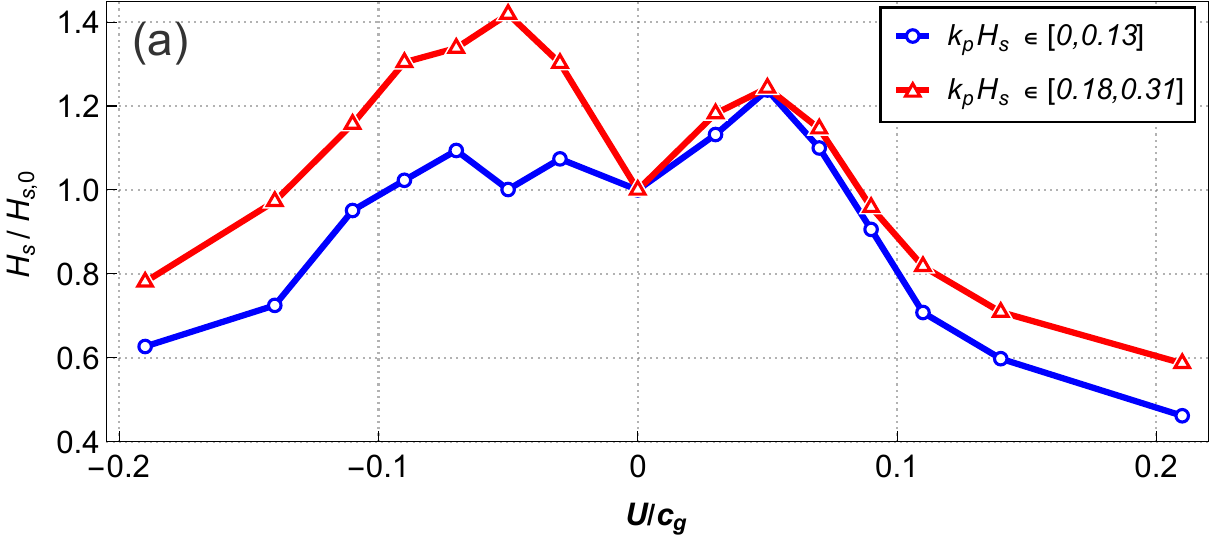}
\endminipage

\hspace{-0.5cm}
\minipage{0.6\textwidth}
    \includegraphics[scale=0.53]{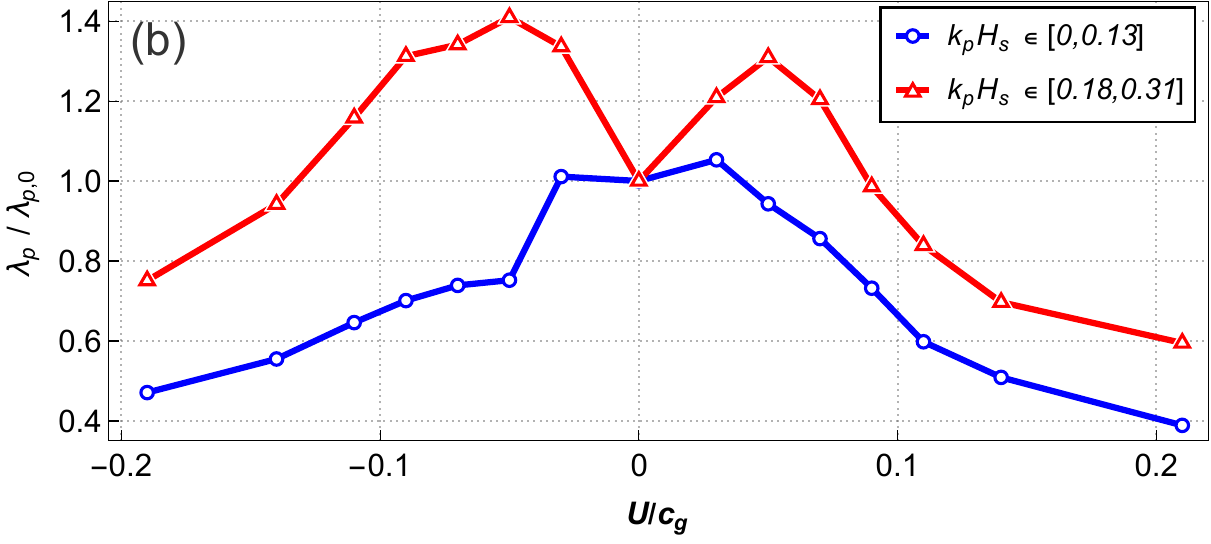}
\endminipage

\hspace{-0.5cm}
\minipage{0.6\textwidth}
    \includegraphics[scale=0.53]{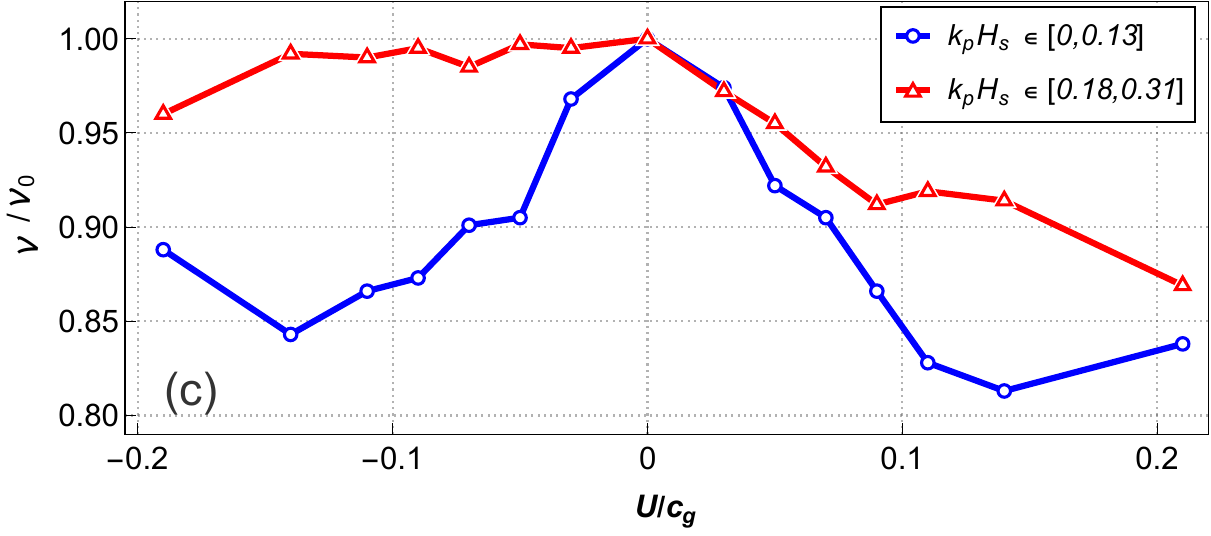}
\endminipage
\caption{Relative change in the average values of significant wave height, wavelength, and bandwidth as a function of normalized tidal stream speed over different partitions of the North Sea data.}
\label{fig:spectraldist3}
\end{figure*}
As expected due to the Doppler effect~\cite{Peregrine1976}, wavelength is downshifted by an opposing current, and upshifted by a slight following current (\jfm{figure} \ref{fig:spectraldist00}\jfm{d}) for mild current speeds. However, over most of the considered range of following current ($U/c_g > 0.03$), the wavelength shifts back down almost symmetrically to the opposing current. Accordingly, this symmetry also appears (although with an inverted shape) in the relative water depth $k_p h$ (\jfm{figure} \ref{fig:spectraldist00}\jfm{c}). From the point of view of the classical treatment of waves interacting with uniform currents, this reversal of the Doppler effect is unexpected \citep{Peregrine1976}. For the spectral bandwidth and directional spread, both parameters show symmetrical and bell-shaped responses to the following and opposing currents (\jfm{figures} \ref{fig:spectraldist00}\jfm{e,f}).

From the classical literature \citep{Peregrine1976}, we know that the dispersion relation of narrow-banded, unidirectional, and linear waves is affected by the presence of a current: The dispersion relation $\omega_0^2 = gk_0$ becomes $(\omega \pm Uk)^2 = gk_0$. Bearing in mind that in deep water $\omega / k = c_p = 2 c_g$, with $c_p$ denoting the phase \x{speed}, one finds the modulation of the wave number to be of the order of,
\begin{equation}
\x{
\frac{k}{k_0} \approx \frac{1}{[1 \pm (U/2c_g)]^2} \approx 1 \mp (U/c_g) + 3/4 (U/c_g)^2 \quad .
}
\end{equation}
When $|U/c_g| \ll 0.1$, this modulation is expected to be strongly asymmetrical, however larger values of $U/c_g \x{= \gamma}$ will diminish such asymmetry, as the quadratic term increases and tends to dominate. \x{A less simplistic computation can be performed for non-homogeneous currents, and \citet{Higgins1960} showed that polynomials with order of $\gamma$ and $\gamma^2$ will appear.} Furthermore, the wave theory for steeper waves leads to a strengthening of the quadratic term~\citep{Brevik1978,Bredmose2001,Jonsson1978}, in addition to shear terms \citep{Swan2000,Swan2001}. Besides, from the dynamic point of view, \citet{Pizzo2023} showed that the relevant parameter is the phase speed of the current relative to the wave group \x{velocity}, so that the interaction between a following current and a comparatively slow wave field in the fixed frame appears to be an interaction between an opposing current and waves in the reference frame of the tide.

To rule out artifacts related to the depth at which the tidal stream speed was considered, 
we compared the previous analysis to its counterpart performed at the middle point of the column ($h$~=~-15~m). These returned very similar results. Note that deeper data typically features 10\% more samples, due to the reduced perturbations of the ADCP signal. Averaging the velocity between 7.5 and 22.5~m depth also provided similar results, depicted in \jfm{figure} \ref{fig:spectraldist0}.

The symmetric behavior of the significant wave height in \jfm{figure} \ref{fig:spectraldist00}\jfm{b} with regard to the direction of the tidal current $U/c_g$ mixes different sea states covering a wide range of nonlinearity (or, equivalently, the wave steepness $\varepsilon$, see \jfm{figure}~\ref{fig:hist1}\jfm{g}). Due to higher-order corrections to the dispersion relation~\citep{Brevik1978,Bredmose2001}, different effects of current velocity on wavelength and bandwidth can be expected for different levels of nonlinearity.  
We partitioned the data into three ranges of wave steepness, in accordance with \citeauthor{LeMehaute1976}'s diagram, considering a similar number of waves in each group.

For a low nonlinearity ($\varepsilon \leq 0.06$), the response of the significant wave height to tidal stream speeds is slightly skewed to the right for $|U/c_g| < 0.05$: a following stream boosts the significant wave height more than an opposing one (\jfm{figure} \ref{fig:spectraldist2x}\jfm{a}). For waves of higher orders in steepness ($\varepsilon > 0.08$) ranging up to breaking, we observed a higher increase of $H_s$ for opposing streams, which is the opposite of what was observed for linear seas (\jfm{figure} \ref{fig:spectraldist2x}\jfm{c}), while intermediate nonlinearities feature an intermediate behavior, with an approximately symmetrical effect of currents on $H_s$ (\jfm{figure} \ref{fig:spectraldist2x}\jfm{b}). These skewnesses are highly significant, the Welch's $t$-test between bins with forward and opposing streams of equal speed displaying $p$ values smaller than $10^{-3}$.

In all cases, the significant wave height was damped by currents, whether opposing or following, beyond $|U/c_g| > 0.05$. Note that the pattern observed in \jfm{figure} \ref{fig:spectraldist2x}\jfm{a} has been observed previously \citep{Vincent1979,Gemmrich2012}. Thus, apparent contradictions in the literature regarding the primacy of opposing or following tides on the modulation of the significant wave height might be due to the analysis of different ranges of wave steepness. 

\jfm{Figure} \ref{fig:spectraldist3} superposes the statistics for wavelength and bandwidth, along with significant wave height, for low- and high-nonlinearity ranges, normalized with regard to their value in the central bin $U/c_{g}=0$. 
\jfm{Figure} \ref{fig:spectraldist3}\jfm{a} shows that the magnitude of the modulation $H_s/H_{s,0}$ by an opposing current, relative to the no-current case, is higher in nonlinear than in linear seas.  

The wavelength modulation $\lambda_p/\lambda_{p,0}$ is symmetric relative to $U/c_{g}$ regardless of the wave steepness (\jfm{figure} \ref{fig:spectraldist3}\jfm{b}). However, the magnitude of this modulation is larger for higher wave steepness. Conversely, the decay of bandwidth with increasing normalized tidal current speed is quite symmetrical for linear seas and asymmetrical for nonlinear seas (\jfm{figure} \ref{fig:spectraldist3}\jfm{c}). In the latter regime, an opposing current hardly affects the bandwidth, while following currents reduce it.

\section{Conclusion}

We assessed symmetries and asymmetries between the effects of opposing and following currents on wave modulation and their dependence on the normalized tidal speed. 
We showed that following currents will modulate the wave field as strongly, and sometimes even stronger, as the opposing currents. This strong effect of a following current is somewhat unexpected by the classical wave theory, when linear waves are subject to interaction with uniform currents~\cite{Pizzo2023}. Furthermore, we showed that both opposing and following currents narrow the spectrum but increase its directional spread. 
Moreover, the asymmetry in the modulation of significant wave height and the Doppler shift between following and opposing currents only meet the classical theoretical expectations at very low normalized current speeds ($|U/c_g| < 0.03$). At higher current speeds, the shape of the modulation is quite symmetrical for all wave parameters except wave steepness. Whether the maximum modulation appears on the opposing or the following current side depends on the nonlinearity of the sea state and the speed of the tidal stream compared to the wave group velocity.

Our main findings corroborate and extend the study of \citet{Pizzo2023} regarding the modulation of wave fields by a following current, by extending them up to deep water conditions and providing multi-year observational data ensuring robust statistics. However, future work entails reproduction of these findings in the laboratory. By isolating the effects of current speed, bandwidth and directional spread, this could support the elaboration of new theoretical understanding of the modification of the dispersion relation by tidal currents.

\section{Acknowledgements}

The measurement data were collected and made freely available by the BSH marine environmental monitoring network (MARNET), the RAVE project (www.rave-offshore.de), the FINO project (www.fino-offshore.de) and cooperation partners of the BSH. The sea state portal was realized by the RAVE project (Research at alpha ventus), which was funded by the Federal Ministry for Economic Affairs and Climate Action on the basis of a resolution of the German Bundestag.

\section{Data availability statement}

The underlying wave buoy and ACDP data are the property of and were made available by then Federal Maritime and Hydrographic Agency, Germany. They can be obtained upon request from these organizations.

\appendix

\bibliography{Maintext}

\end{document}